\documentclass[aps,apl,preprint,groupedaddress,showpacs]{revtex4}

\usepackage{graphicx}
\begin{document}
\title{Transition on the entropic elasticity of DNA induced by intercalating molecules}
\author{M. S. Rocha}
\author{M. C. Ferreira}
\author{O. N. Mesquita}
\affiliation{{Departamento de F\'\i sica, ICEX, Universidade Federal
de Minas Gerais, \\ Caixa Postal 702, Belo Horizonte, CEP 31270-901,
MG, Brazil}}

\date{\today}

\begin{abstract}
We use optical tweezers to perform stretching experiments on DNA
molecules when interacting with the drugs daunomycin and ethidium
bromide, which intercalate the DNA molecule. These experiments are
performed in the low-force regime from zero up to 2 pN. Our results
show that the persistence length of the DNA-drug complexes increases
strongly as the drug concentration increases up to some critical
value. Above this critical value, the persistence length decreases
abruptly and remains practically constant for larger drug
concentrations. The contour length of the molecules increases
monotonically and saturates as drugs concentration increases.
Measured intercalants critical concentrations for the persistence
length transition coincide with reported values for the helix-coil
transition of DNA-drug complexes, obtained from sedimentation
experiments.


\end{abstract}
 \maketitle

\section{Introduction}\label{s1}

DNA-drug interactions have been much studied along the past years.
An important motivation for these studies is the fact that many of
the studied drugs are used for treatment of human diseases,
particularly, in cancer chemotherapy.

Single molecule stretching experiments using optical tweezers have
yielded a great amount of information about DNA-protein and DNA-drug
interactions \cite{Ashkin1, Smith, Wang, Shivashankar2, Baumann,
Grier}.

Recently, we studied the interaction between psoralen and DNA when
illuminated with ultraviolet light A (UVA). Psoralen is a drug used
to treat some skin diseases, like psoriasis and vitiligo. This drug
intercalates the DNA molecule and can form covalent linkages with
the thymines if the complex is illuminated with ultraviolet light,
modifying drastically its elasticity and impeding the DNA
replication and transcription. The persistence length of the
DNA-psoralen complexes formed after UVA illumination were measured
in \cite{Rocha}.

Daunomycin and ethidium bromide (EtBr) are other examples of drugs
which intercalate the DNA molecule and can modify its elasticity,
depending on the drug concentration. Both drugs unwind the DNA
double helix when intercalating \cite{Fritzsche}. Daunomycin is an
anthracycline antibiotic used in the treatment of various cancers.
It inhibits DNA replication and transcription when intercalating,
impeding cell duplication \cite{Chaires}. Ethidium bromide (EtBr) is
commonly used as a non-radioactive marker for identifying and
visualizing nucleic acid bands in electrophoresis and in other
methods of nucleic acid separation.

Several works have reported different results for the effects of
these drugs on the entropic elasticity of DNA molecules. In those
works, the measured parameter used to study elasticity modifications
is the persistence length of the DNA-drug complex. Smith \textit{et
al.} \cite{Smith} report that ethidium bromide does not modify the
elasticity of the DNA molecule, only increasing its contour length
by $\sim$40$\%$. Tessmer \textit{et al.} \cite{Tessmer} report that
ethidium bromide causes a large increase in the contour length and a
decrease in the persistence length of the complex for 1 $\mu$M of
the drug, and at lower concentrations, an increase in both
persistence and contour lengths. Recently, Sischka $\textit{et al.}$
\cite{Sischka} report the value of 28.1 nm for the persistence
length of DNA-daunomycin complexes and 20.7 nm for DNA-EtBr
complexes, smaller than the bare DNA persistence length of about 50
nm. The authors have used in this work a concentration of 1 $\mu$M
for both drugs, and a DNA concentration of 15 pM. In the present
work, in order to clearly establish the effect of these
intercalating drugs on the persistence length of the DNA complexes,
we performed stretching experiments at various drugs concentrations,
from zero up to saturation of the complexes. We show that the values
obtained for the persistence length depend strongly on the
concentration ratio between drug and DNA base pairs. Our results
show that the persistence length of the complexes increases as we
increase the drug concentration until certain critical concentration
is reached. Above this critical concentration the persistence length
decreases abruptly and remains practically constant for larger drug
concentrations.

\section{Experimental procedure}\label{s2}

To measure the persistence and contour length of DNA molecules and
DNA-drug complexes, we use optical tweezers \cite{Ashkin1, Smith,
Wang, Shivashankar2, Baumann, Grier} and intensity autocorrelation
spectroscopy \cite{Nathan}.

The samples consist of $\lambda$-DNA molecules in a PBS pH 7.4 with
[NaCl] = 140 mM solution. We attach one end of the molecule to a
microscope coverslip, and the other end is attached to a polystyrene
bead. To do this, we use the procedure described in
\cite{Shivashankar}. We add the drug in the sample immediately
before the measurements. The DNA concentration used in all
experiments was $C_{DNA}$ = 6.81 $\mu$g/mL, which corresponds to a
base pairs concentration of $C_{bp}$ = 11 $\mu$M.

Our optical tweezers is mounted in a Nikon TE300 microscope with an
infinite corrected objective (100X, N.A. = 1.4). The trapping laser
is an infrared (IR) laser with $\lambda$ = 832 nm (SDL, 5422-H1).
The optical tweezers is used to trap the polystyrene bead attached
to the end of the DNA molecule, so we can manipulate and stretch the
DNA molecule.

In addition, we use a He-Ne laser ($\lambda$ = 632.8 nm) as the
scattering probe. The backscattered light by the polystyrene bead is
collected by a photodetector, which delivers pulses to a digital
correlator. We, then, obtain the autocorrelation function of the
backscattered light, which allows us to determine the stiffness of
the optical trap, due to the Brownian motion of the trapped bead.

The next step in the experimental procedure is to obtain the force
\textit{versus} extension curves for the DNA molecules and DNA-drug
complexes. To do this, we use the optical tweezers to trap the bead
with the DNA while pulling the microscope slide, stretching the DNA.
The backscattered light is collected while stretching the DNA. From
the backscattered light intensity one obtains the displacement of
the trapped bead in relation to its equilibrium position, and by
multiplying it times the tweezers' stiffness, the force exerted by
the DNA molecule while it is stretched is obtained. The details
about our experimental setup and experimental procedure can be found
in \cite{Rocha, Nathan}.

Finally, with the force \textit{versus} extension curves, we use the
approximate expression derived by Marko and Siggia (Eq.
\ref{markosiggia}) \cite{Marko} to obtain the persistence and
contour length of the DNA molecules and DNA-drug complexes,

\begin{equation}
F = \frac{k_BT}{A}\left[\frac{z}{L} + \frac{1}{4\left(1 -
\frac{z}{L}\right)^2} - \frac{1}{4}\right] \label{markosiggia},
\end{equation}
where $z$ is the DNA molecule end-to-end distance, $\textit{k$_B$}$
is the Boltzmann constant, $\textit{T}$ is the absolute temperature,
$\textit{A}$ is the DNA persistence length and $\textit{L}$ is the
DNA contour length.

\begin{figure}
\begin{center}
\includegraphics[width=8cm]{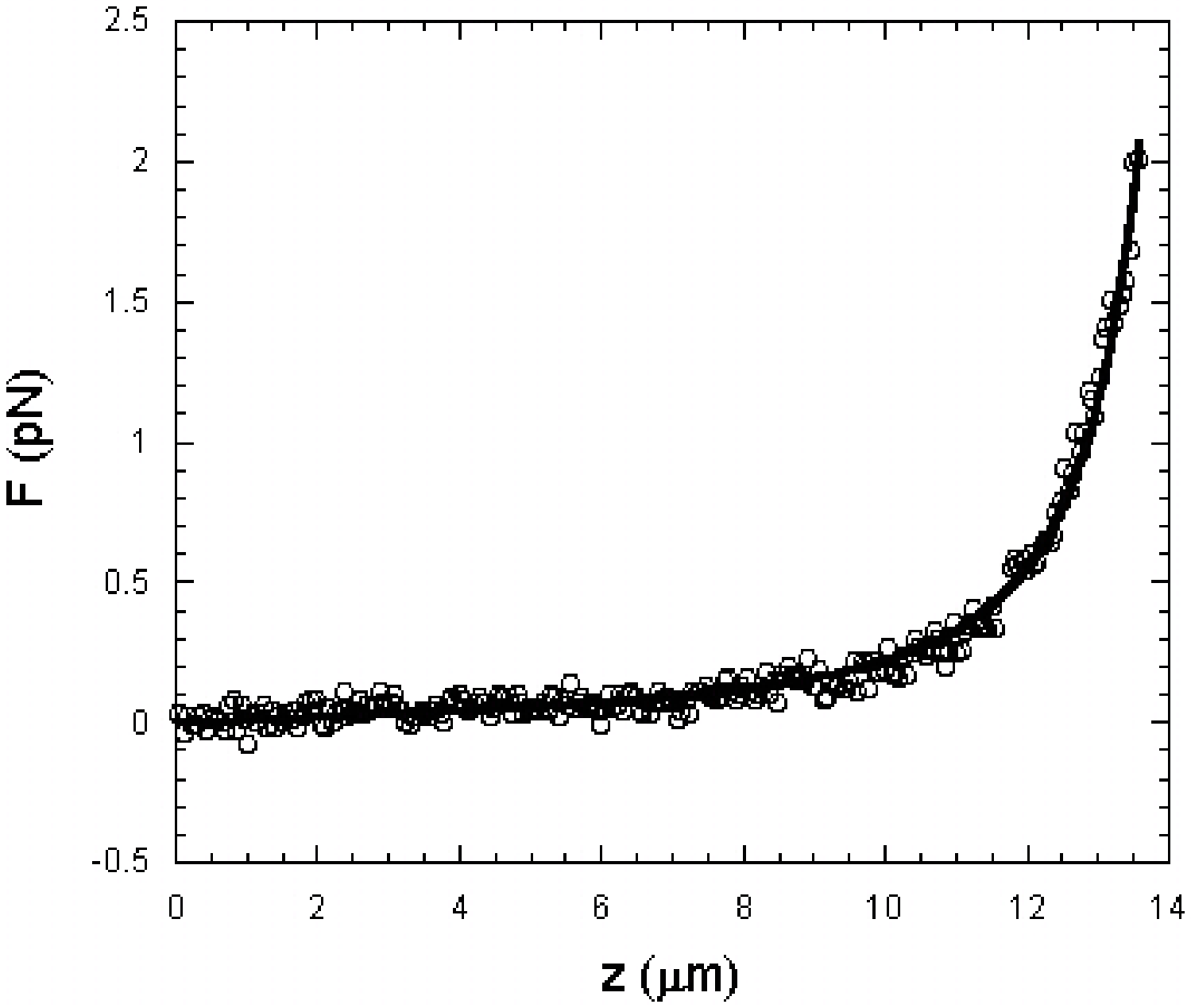}
\end{center}
\caption{} \label{force}
\end{figure}

Figure \ref{force} is a typical force \textit{versus} extension
curve obtained with this procedure, for a drug-free DNA molecule.
Fitting this curve with Eq. \ref{markosiggia}, we extract the
persistence length ($A$ = 50 $\pm$ 3 nm) and the contour length ($L$
= 16.5 $\pm$ 1 $\mu$m) for the $\lambda$-DNA. These values
correspond to well-known values reported in the literature
\cite{Strick2, Wang, Coelho}.

\section{Results and discussion}\label{s3}

In this section we show the results obtained for the two drugs used:
daunomycin and ethidium bromide.

\subsection{Daunomycin}\label{s31}

We have performed experiments with DNA-daunomycin complexes for
several drug concentrations. In Fig. \ref{daunomycin} we show the
persistence length ($A$) of the complexes as a function of total
daunomycin concentration ($C_D$) for fixed DNA base pairs
concentration of $C_{bp}$ = 11 $\mu$M. We denote by $C_D$ the total
daunomycin concentration used to prepare the sample, which is the
sum of both the bounded to DNA and the free drug concentration in
solution.

The point which the drug concentration is zero corresponds to the
drug-free DNA situation with $A$ = 50 nm.

The behavior of the persistence length $A$ as a function of
daunomycin concentration $C_D$ can be described as follows: it
initially increases with $C_D$ until it reaches a maximum value
($\sim$ 280 nm) at the critical concentration $C_D^{critical}$ =
18.3 $\mu$M. Then, it decays abruptly to around 75 nm and remains
practically constant at this value even if we continue to increase
$C_D$. The contour length increases monotonically from 16.5 $\pm$ 1
$\mu$m up to the saturation value of 21 $\pm$ 1 $\mu$m. These mean
values are obtained performing an average over many different DNA
molecules and DNA-drug complexes. Such distribution of contour
length values was also observed by Mihailovic \textit{et al.}
\cite{Mihailovic}.

Also, we can estimate the exclusion parameter $n$ (number of total
base pairs divided by the number of total intercalated drug
molecules) from our experimental data. The average value of the
contour length for DNA-daunomycin complexes obtained when using a
saturated concentration of the drug increases about 27$\%$ relative
to drug-free DNA contour length (16.5 $\mu$m). This means that when
all possible drug molecules are intercalated, the DNA increases its
contour length by 4.5 $\mu$m. Knowing that each intercalated
daunomycin molecule increases the contour length of the complex by
0.31 nm \cite{Fritzsche}, we determine the total number of
intercalated daunomycin molecules, which is around 14500. Finally,
the exclusion parameter can be obtained by dividing the number of
base pairs of the $\lambda$-DNA (48500) by the number of total
intercalated drug molecules (14500). We obtain $n$ = 3.3, in good
agreement with the value 3.5 reported in \cite{Chaires}.

\begin{figure}
\begin{center}
\includegraphics[scale=.5]{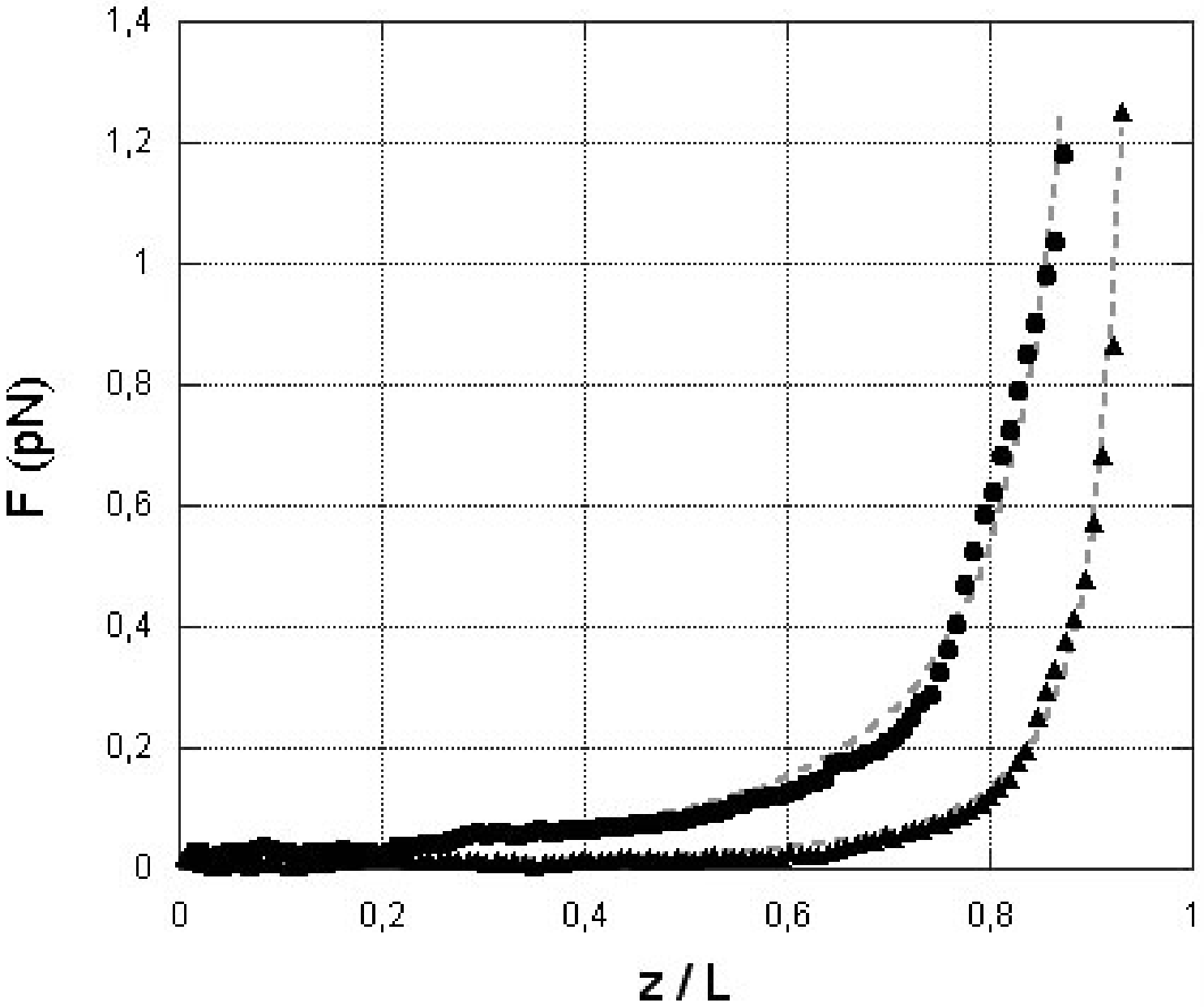}
\end{center}
\caption{} \label{Fdouble}
\end{figure}

For comparison purposes, Fig. \ref{Fdouble} shows two force
\textit{versus} extension curves (normalized by the contour length)
for two daunomycin concentrations, before and after the transition.
The data points in this figure are smoothed for better
visualization, \textit{i. e.}, Brownian fluctuations are averaged
out.

\subsection{Ethidium Bromide (EtBr)}\label{s32}

The behavior of the persistence length as a function of the drug
concentration for DNA-EtBr complexes is very similar to the
DNA-daunomycin complexes. The difference is that in this case the
transition occurs at a lower drug concentration (see Fig.
\ref{bromide}) for the same DNA base pairs concentration $C_{bp}$ =
11 $\mu$M. The maximum value measured for the persistence length of
DNA-EtBr complexes is $\sim$ 150 nm, at the critical concentration
$C_E^{critical}$ = 3.1 $\mu$M. The contour length increases
monotonically from 16.5 $\pm$ 1 $\mu$m up to the saturation value of
23 $\pm$ 1 $\mu$m. Again, these values for the contour lengths are
averages over many different molecules.

Repeating the same calculation for the exclusion parameter of EtBr,
which increases the DNA contour length by 0.34 nm per intercalated
molecule \cite{Sischka}, we obtain $n$ = 2.5, in reasonable
agreement with the value 2.01 reported in \cite{Coury}.

\subsection{Equilibrium binding constants}\label{s34}

In our experiments we control the total drug concentration $C_T$ and
the total concentration of DNA base pairs $C_{bp}$. To discuss the
elastic properties of the DNA complex formed the important parameter
to consider is the ratio $r$ between the concentration of bounded
drug ($C_b$) per concentration of DNA base pairs ($C_{bp}$). In
order to obtain $r$, the binding of molecules to DNA is analyzed
using the neighbor exclusion model \cite{Chaires}. A closed form for
this model was obtained by McGhee and von Hippel \cite{McGhee} and
can be expressed by the equation

\begin{equation}
\frac{r}{C_f} = K_i (1 - nr) \left[\frac{1 - nr}{1 - (n - 1)r
}\right]^{n - 1}  \label{nem},
\end{equation}
where $r$ is ratio between the concentration of bounded drug ($C_b$)
per concentration of DNA base pairs ($C_{bp}$), $C_f$ is the
concentration of free drug (not bounded), $K_i$ is the intrinsic
binding constant and $n$ is the exclusion parameter in base pairs.
For a more detailed discussion about the neighbor exclusion model,
see \cite{McGhee}.

The concentration of free drug ($C_f$) can be simply related with
the concentration of bounded drug ($C_b$) and the total drug
concentration ($C_T$) through the equation

\begin{equation}
C_T = C_f + C_b \label{ct}.
\end{equation}

Using Eq. \ref{nem} and \ref{ct} with the determined exclusion
parameter ($n$ = 3.3), the intrinsic binding constant reported in
\cite{Chaires} for daunomycin, $K_i$ = 7$\times$10$^5$ M$^{-1}$, the
critical daunomycin concentration measured in this work
($C_D^{critical}$ = 18.3 $\mu$M), and the concentration of DNA base
pair used in our experiments ($C_{bp}$ = 11 $\mu$M), we can
determine the critical ratio $r_c$, which we define as the value of
$r$ at the abrupt transition for the value of the persistence
length. We then obtain the value $r_c$ = 0.248.

Similarly, for EtBr, we use the parameters $n$ = 2.5, $K_i$ =
1.5$\times$10$^5$ M$^{-1}$ \cite{Gaugain}, and $C_E^{critical}$ =
3.1 $\mu$M determined again from the abrupt change in persistence
length. We obtain $r_c$ = 0.131. In Section \ref{s35} we compare the
values obtained for $r_c$ with those reported in the literature for
a sedimentation experiment.

It is important to mention that $K_i$ varies with the ionic strength
of the solution. The values used here are the values for the ionic
concentrations used in our experiments.

\subsection{Interpretation of the DNA-drug complexes elasticity results}\label{s35}

\begin{figure}
\begin{center}
\includegraphics[scale=.5]{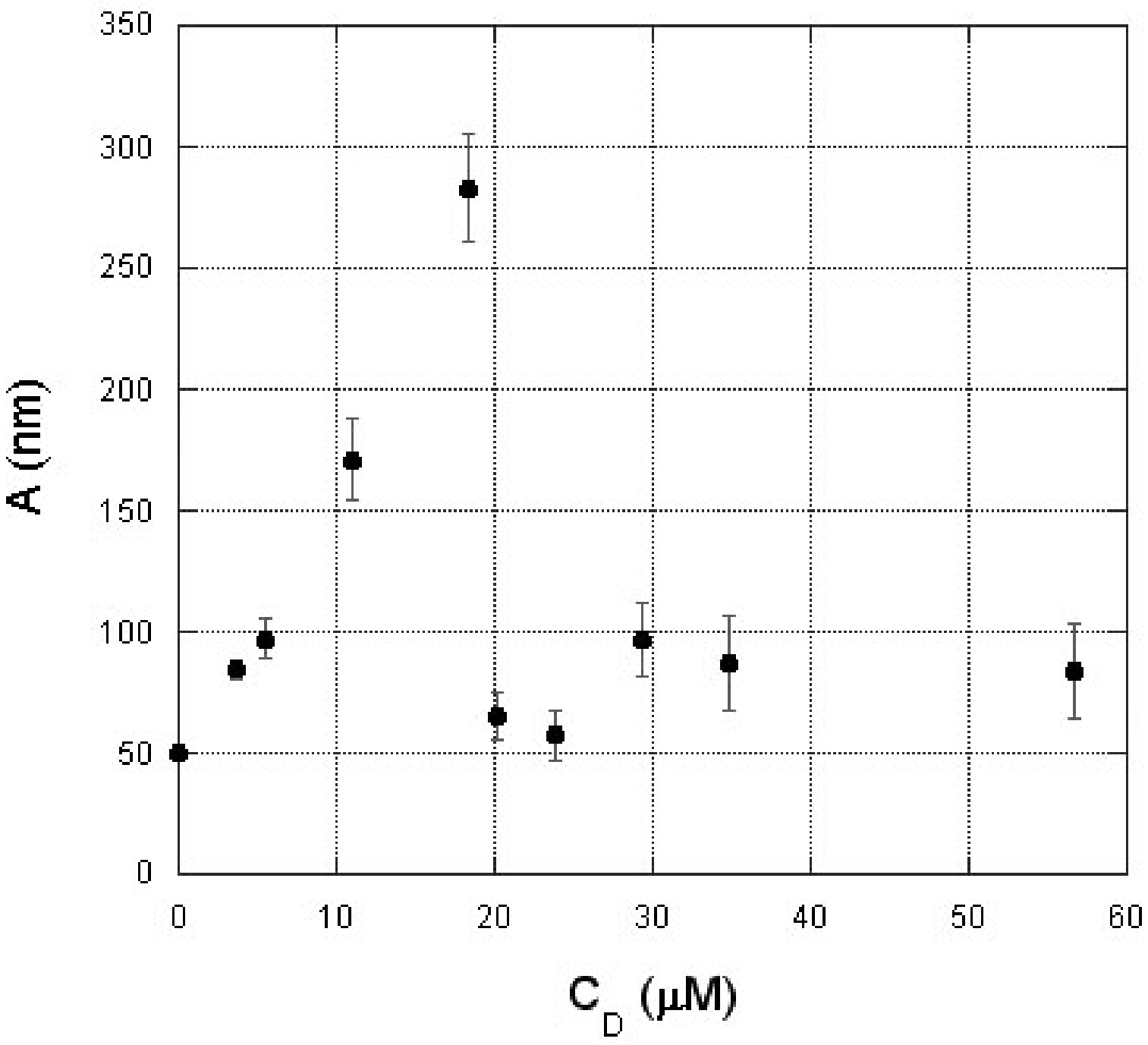}
\end{center}
\caption{} \label{daunomycin}
\end{figure}

For low drug concentrations, drug intercalation in the DNA molecule
increases the rigidity of the complex (see Figs. \ref{daunomycin}
and \ref{bromide}). This is consistent with the results of Vladescu
\textit{et al.} \cite{Vladescu}, which shows that EtBr stabilizes
the DNA double-helix for low drug concentrations. They have
performed melting experiments with various EtBr concentrations, from
zero to 2.5 $\mu$M, showing that EtBr intercalation stabilizes the
DNA double-helix in this concentration range. Therefore, we expect
an increase of the persistence length of DNA-drug complexes in this
low concentration range. Figure \ref{bromide} shows this increase
for EtBr, and Fig. \ref{daunomycin} shows a similar result for
daunomycin.

For high drug concentrations, \textit{i. e.}, above the critical
concentration (peak of Figs. \ref{daunomycin} and \ref{bromide}),
the persistence length of the complexes decays abruptly and remains
practically constant.

It is well-known that intercalation unwinds the DNA double-helix
\cite{Fritzsche}. Due to unwinding and above some drug critical
concentration, the complexes can have a helix-coil transition, which
can cause DNA denaturing as the DNA is stretched, decreasing the
persistence length of DNA-drug complexes as seen in Figs.
\ref{daunomycin} and \ref{bromide}. The unwinding angle per
intercalated EtBr drug molecule is approximately 1.7 times greater
than that for daunomycin intercalation \cite{Fritzsche}. Therefore
we expect that the transition occurs for EtBr at a lower drug
concentration as compared with daunomycin, if the same DNA
concentration is used. This is confirmed experimentally in our data
of Figs. \ref{daunomycin} and \ref{bromide}.

Sedimentation experiments performed with circular DNA as a function
of daunomycin and ethidium bromide concentrations display a minimum
in the sedimentation coefficient S$_{20}$ at $r_c$ = 0.192, for
daunomycin and $r_c$ = 0.114 for ethidium bromide \cite{Fritzsche}.
The minimum in the sedimentation coefficient S$_{20}$ is associated
with a helix-coil transition, due to unwinding of the DNA
double-helix by the intercalating drugs \cite{Fritzsche}. These
numbers agree within 15 to 30$\%$ with the values of $r_c$
determined from our DNA persistence length measurements. This
indicates that the abrupt change of the DNA persistence length for
both drugs might be also caused by a helix-coil transition due to
the unwinding of the DNA double-helix as the drugs intercalate into
it. The reasonable agreement between the critical ratios $r_c$
obtained from sedimentation experiments \cite{Fritzsche} and from
our measurements of the persistence length transition provides an
evidence that a helix-coil transition is probably what we are
observing in our experiments.

In addition, it is known that EtBr (and also most intercalating
drugs) exhibits multimodality at their interaction with DNA
\cite{Vardevanyan, Arabzadeh}. The type of interaction varies with
the drug concentration. The abrupt transition shown in Figs.
\ref{daunomycin} and \ref{bromide} might as well be caused by
different ways of drug binding to DNA.

\section{Conclusion}\label{s4}

We have made systematic measurements of the entropic elasticity
variation of a $\lambda$-DNA molecule when interacting with two
drugs, daunomycin and ethidium bromide, as a function of their
concentrations. Our results show that the persistence length of the
DNA-drug complexes increases strongly as the drug concentration
increases, for low concentrations. Above certain critical drug
concentration the persistence length decreases abruptly and remains
practically constant for high drug concentrations. This behavior is
quite similar for both daunomycin and EtBr, as shown in Figs.
\ref{daunomycin} and \ref{bromide}. Our results suggests that the
abrupt transition observed in the persistence length might be due to
a helix-coil transition and denaturing of DNA-drug complexes above
the critical concentration, resulting in a decrease of the
persistence length.

\section{Acknowledgements} \label{s5}

This work was supported by the Brazilian agencies: Conselho Nacional
de Desenvolvimento Cient\'ifico e Tecnol\'ogico (CNPq), FAPEMIG,
FINEP-PRONEX, Instituto do Mil\^enio de Nanotecnologia e Instituto
do Mil\^enio de \'Optica N\~ao-linear e Biofot\^onica - MCT. M. S.
R. acknowledges support by LNLS.

\bibliography{rocha}

\clearpage
\section*{Figure Legends}
\subsubsection*{Figure~\ref{force}.}
Force as a function of extension for a drug-free DNA molecule. By
fitting this curve with Eq. \ref{markosiggia}, we determine the
persistence length $\textit{A}$ = 50 $\pm$ 3 nm and the contour
length $\textit{L}$ = 16.5 $\pm$ 1 $\mu$m.

\subsubsection*{Figure~\ref{daunomycin}.}
Persistence length $A$ of DNA-daunomycin complexes as a function of
drug concentration for fixed DNA concentration ($C_{bp}$ = 11
$\mu$M). $A$ initially increases with $C_D$ until it reaches a
maximum value ($\sim$ 280 nm) at the critical concentration
$C_D^{critical}$ = 18.3 $\mu$M. Then, the persistence length decays
abruptly to around 75 nm and remains practically constant at this
value even if we continue to increase the drug concentration.

\subsubsection*{Figure~\ref{Fdouble}.}
Force \textit{versus} extension curves (normalized by the contour
length) for two daunomycin concentrations. The data Brownian
fluctuations are averaged out for better visualization.
\textit{Circles}: $C_D$ = 20.1 $\mu$M (above the critical
concentration) and $A$ $\sim$ 61 nm; \textit{triangles}: $C_D$ =
18.3 $\mu$M and $A$ $\sim$ 263 nm. Dashed lines are fittings using
Eq. \ref{markosiggia}.

\begin{figure}
\begin{center}
\includegraphics[scale=.5]{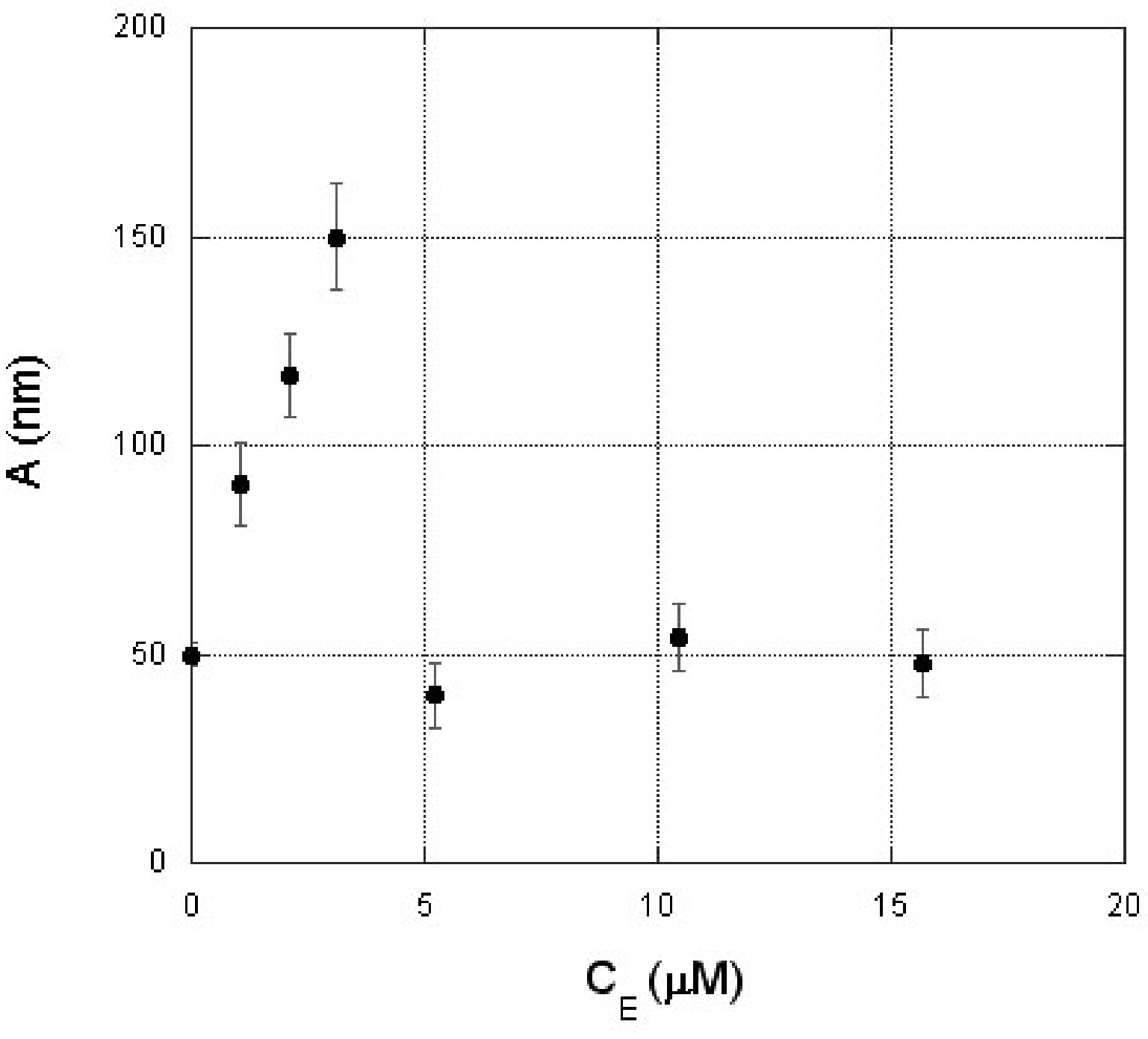}
\end{center}
\caption{} \label{bromide}
\end{figure}

\subsubsection*{Figure~\ref{bromide}.}
Persistence length of DNA-EtBr complexes as a function of drug
concentration for fixed DNA concentration ($C_{bp}$ = 11 $\mu$M).
Here, the transition occurs at a lower drug concentration. The
maximum value measured for the persistence length of DNA-EtBr
complexes is $\sim$ 150 nm, at the critical concentration
$C_E^{critical}$ = 3.1 $\mu$M.

\clearpage

\clearpage

\clearpage

\clearpage

\end{document}